\documentclass[pre,aps,floatfix,notitlepage,10pt,twocolumn]{revtex4-1}
\usepackage{amsmath}
\usepackage{amssymb,amsfonts,latexsym}
\usepackage{bm}
\usepackage[mathcal]{euscript}
\usepackage[dvips]{graphicx}
\usepackage{epsfig}
\usepackage{color}
\usepackage{xspace}
\usepackage{grfext}

\PrependGraphicsExtensions*{.pdf,.PDF}  %Looks for PDF figures first in case there are also PNG
\newcommand{\ie}{\textit{i.e.}\xspace}
\newcommand{\eg}{\textit{e.g.}\xspace}
\newcommand{\etal}{et al.\xspace}
\newcommand{\Dr}{D_\text{r}}
\newcommand{\vp}{v_\text{p}}
\def\rhog{\rho_\text{g}}

\begin{document}
\title{Emergent Self-organization in Active Materials}

\author{Michael F. Hagan}\email{hagan@brandeis.edu}
\author{Aparna Baskaran}\email{aparna@brandeis.edu}
\affiliation{Department of Physics, Brandeis University, Waltham MA, 02454, U.S.A.}

\date{\today}

\begin{abstract}
Biological systems exhibit large-scale self-organized dynamics and structures which enable organisms to perform the functions of life. The field of active matter strives to develop and understand microscopically-driven nonequilibrium materials, with emergent properties comparable to those of living systems. This review will describe two recently developed classes of active matter systems, in which simple building blocks --- self-propelled colloidal particles or extensile rod-like particles --- self-organize to form macroscopic structures with features not possible in equilibrium systems. We summarize the recent experimental and theoretical progress on each of these systems, and we present simple descriptions of the physics underlying their emergent behaviors.
\end{abstract}

%\begin{keyword}
%active matter \sep active nematic \sep simulation \sep hydrodynamic theory \sep self-propelled colloid \sep self-organization \sep non-equilibrium
%\end{keyword}

\maketitle

\section*{Introduction}
A biological cell is a marvel of natural engineering. Within its cytoplasm, proteins collectively self-organize into dynamical assemblies which enable the cell to crawl or swim, replicate, and self-heal. In comparison, traditional human-engineered materials have remarkably limited functionality, since they are constrained by the laws of thermodynamics which forbid emergence of spontaneous motion or macroscopic work. The cell is freed from the rigid constraints of equilibrium because its biomolecules are `active', converting energy into motion at the microscale, to maintain the cell out of equilibrium. The paradigm of \textit{active matter} encompasses materials whose constituent elements, like the components of a cell, consume energy to generate forces and motion.

The most plentiful source of active materials is biology itself, where active matter is found on scales ranging from the cell cytoskeleton to swarming bacterial colonies and flocking animals. Further, it is now possible to construct artificial `active materials' from biomolecules or synthetic colloids which have a similar capability to convert energy into motion. These systems are poised to transform materials science, creating self-organizing materials capable of autonomous motion or adapting to their environments, thus mimicking the capabilities of living organisms. However, there is currently no predictive theory that enables designing components to self-organize into particular structures or perform particular functions. Thus, a key goal of active matter is to develop such a theoretical framework, by starting from controllable experimental model systems and minimal computational models, and identifying relevant symmetry principles and conservation laws. In this review, we summarize recent progress toward this goal, arising from two recently developed model systems.

\begin{figure*} [hbt]
\centering{\includegraphics[width=0.95\textwidth]{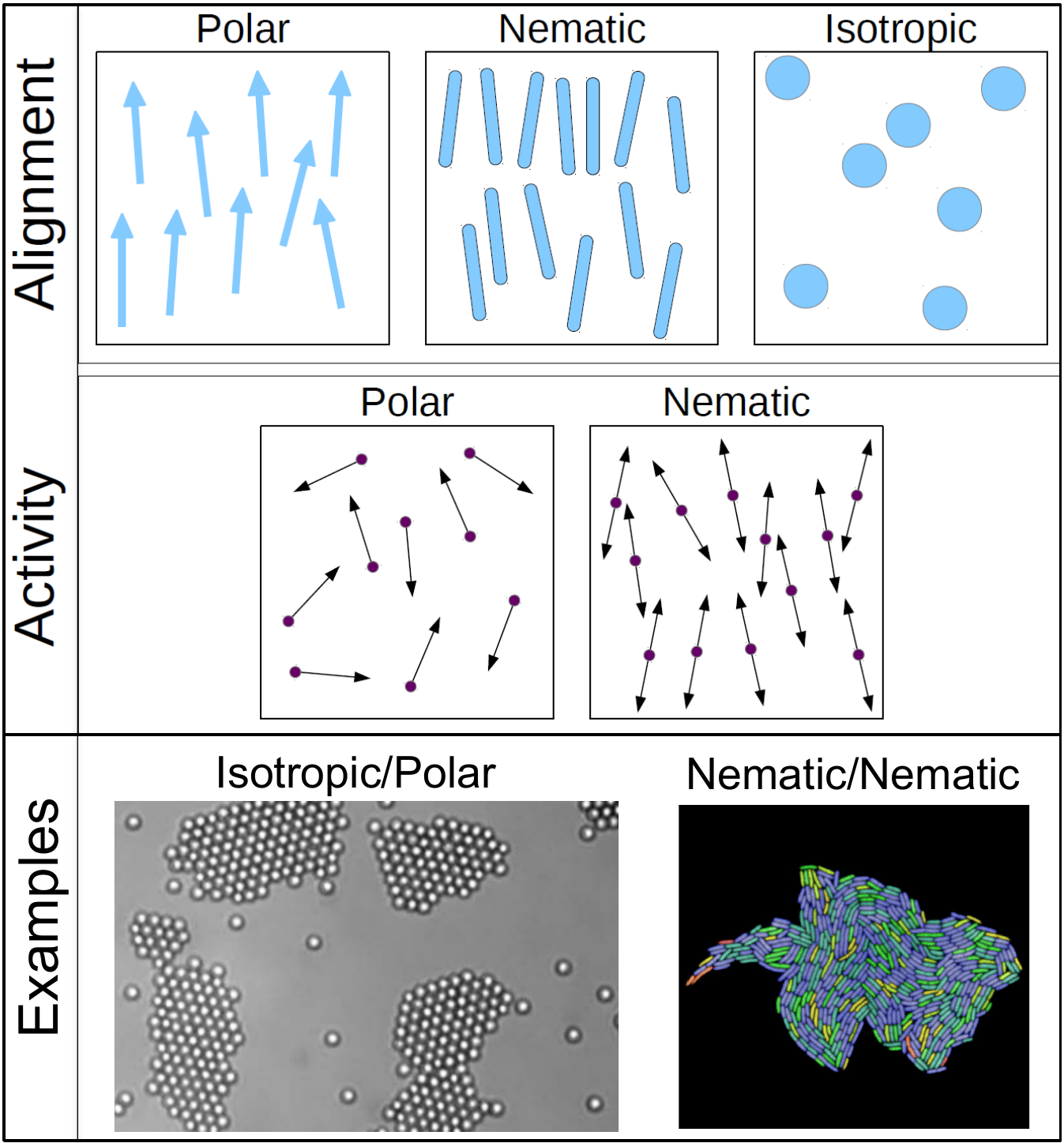}}
\caption{Active systems can be categorized by two types of symmetries: the symmetry of
alignment between particles (top row), and the symmetry of the activity (middle row). Only the most common symmetries are shown for brevity:  particles which align along a preferred direction (polar), particles which align along a preferred axis, but with head-tail symmetry (nematic), or particles which have no preferred direction of alignment (isotropic). Some examples of active systems identified according to their symmetries are in the bottom row : Left: Self-propelled Janus colloids that have a polar activity but isotropic interactions (Image Credit : Jeremi Palacci \cite{Palacci2013a}); Right: A dividing colony of rod-like bacteria that has nematic activity due to the growth at both ends and nematic interactions due to their shape (Image from \cite{EColiImage2005}).
}
\label{fig:Symmetries}
\end{figure*}

A reasonable starting point to classify active matter systems is based on two symmetries --- one associated with how particles align with each other, and another associated with how activity enters the system. The first is familiar from equilibrium materials. Particles with isotropic interactions (\eg spherical colloids) form isotropic fluids at moderate densities, whereas elongated particles align to form orientationally ordered liquid crystals (such as nematic liquid crystals used in displays). In contrast, the symmetry of the activity only applies to nonequilibrium, active systems. For example, a bird flying in the direction of its head undergoes directed motion and has polar activity, whereas a bacterium growing from both ends has nematic activity. As illustrated in Fig. [\ref{fig:Symmetries}], these symmetries can be combined in different ways to generate different classes of behaviors.

Motivated by flocks of animals, the first- and most-studied symmetry class is polar alignment and polar activity, or particles that align in the same direction as their neighbors and undergo head-directed motion, generating polar flocks. Early theoretical work showed that hydrodynamic symmetry-based models can describe the macroscopic behaviors of such flocks, and demonstrated that active matter systems can exhibit behaviors not possible in equilibrium systems, such as long-range order in 2D \cite{Toner1995}. More recently, experimentalists have created active matter systems with other kinds of symmetries, leading to qualitatively new kinds of behaviors. In this review, we focus on two of these systems, where at first glance we might not expect coherent emergent behaviors:  particles with polar activity but no interparticle alignment, and particles that align but experience nematic activity. In each case, we summarize experiments and modeling that discovered new ordered phases, and we identify the physics underlying these phases.

\def\Pe{\text{Pe}}
\begin{figure*} [hbt]
\centering{\includegraphics[width=0.95\textwidth]{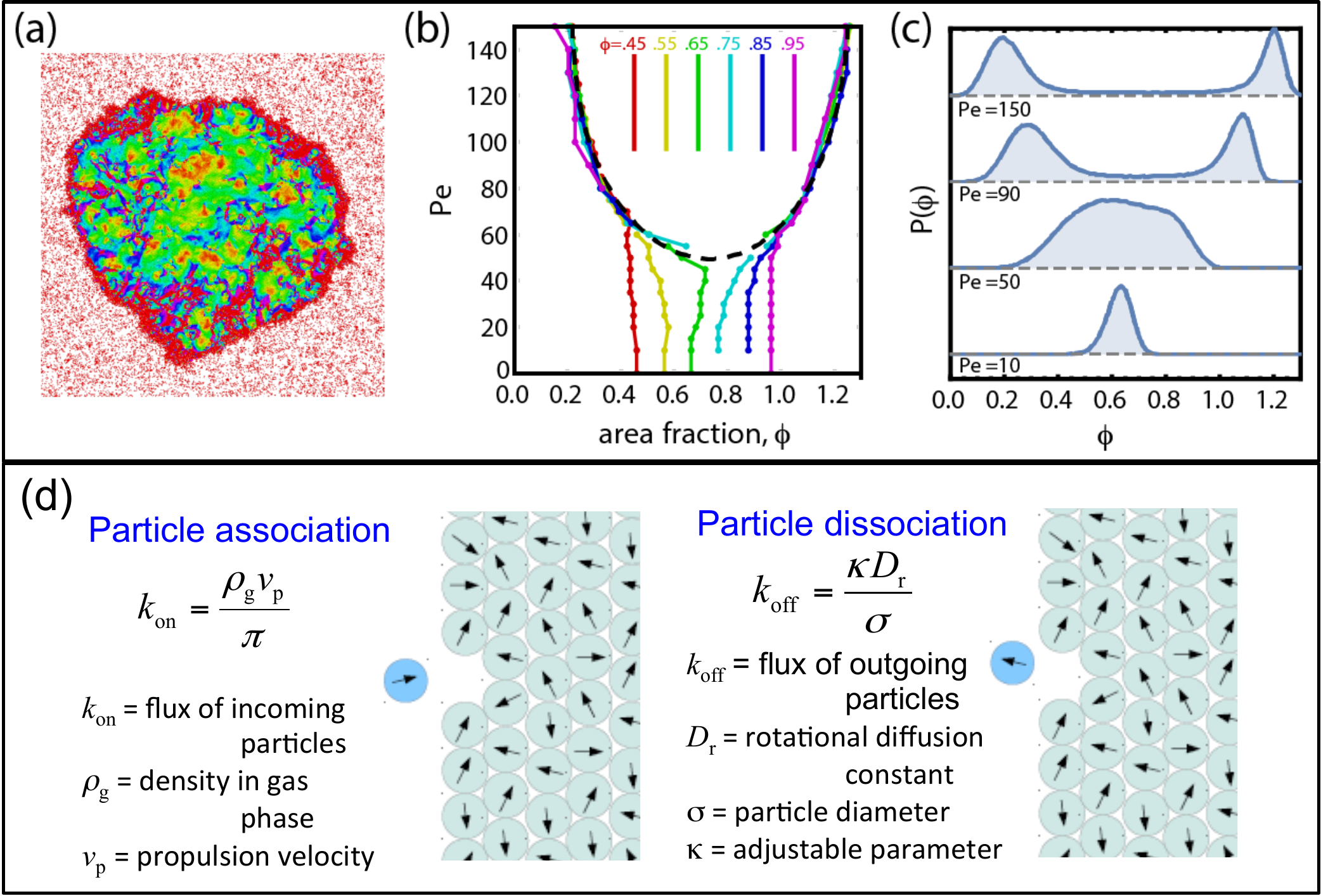}}
\caption{Motility-induced phase separation of self-propelled spheres. \textbf{(a)} Snapshot from a simulation of self-propelled spheres in phase coexistence. The color describes particle speed (increasing from blue to red), illustrating the inhomogeneous motion within the dense phase.
  \textbf{(b)} Phase diagram of the self-propelled sphere system. The area fraction within each phase is shown as a function of P\'eclet number ($\Pe\cong\vp/\Dr$, with $\vp$ the propulsion velocity and $\Dr$ the rotational diffusion constant) for a range of overall area fractions, $\phi$.  At low $\Pe$  the system is single-phase, while above a threshold $\Pe$ it phase-separates. The envelope of coexisting densities is analogous to the binodal of an equilibrium fluid, with $\Pe$ playing the role of an inverse temperature. \textbf{(c)} Distributions of area fractions (measured over small regions within a simulation box) for several values of $\Pe$. In the single-phase regime, below $\Pe \approx 50$, there is a single peak about the overall area fraction (in this case $\phi = 0.65$, near its critical value). The peak broadens and flattens as the critical point is approached, and the distribution becomes bimodal upon entering the phase-separation regime. \textbf{(d)}  Kinetic theory for self-propelled phase separation \cite{Redner2013,Redner2013a}. The flux of self-propelled particles from the dilute phase impinging on surface of a large cluster is  $k_\text{on}=\rhog\vp/\pi$. A particle cannot escape from the surface until its orientation diffuses far enough that it points away from the cluster horizon, so the outgoing flux of particles is given by $k_\text{off}=\kappa \Dr/\sigma$, where $\kappa$ is an adjustable parameter that accounts for the escape of multiple particles at a time among other effects. Calculating the net flux shows that there is a threshold density $\rhog \sim \Dr/\vp$ above which the cluster phase will grow.
}
\label{fig:ActiveSpheres}
\end{figure*}

\section*{Active colloids}
We begin with the simplest model system in materials physics, hard sphere colloids. These are spherical, micron-sized particles with purely repulsive interactions arising from excluded volume. The particles can be suspended in a fluid (3D) or sedimented at an interface (2D), and they form an isotropic gas at most densities. The active analog of a hard sphere is a `self-propelled' colloid that undergoes propulsion along a particular body axis. The closest experimental realization is a spherical particle that is asymmetrically coated with a catalyst (\eg platinum) and placed in solvent with a fuel (\eg hydrogen peroxide). The on-board catalysis generates asymmetric concentrations of products, driving particle motion in a direction determined by its orientation \cite{Howse2007, Ke2010,Theurkauff2012a, Palacci2013a, Palacci2014,Buttinoni2013a}. Further, this orientation fluctuates due to thermal and athermal effects from the medium and hence undergoes rotational diffusion, thereby making the self-propelled colloid perform a persistent random walk. Such a motion is characterized by a \textit{persistence length} (the average distance traveled by a particle before reorienting), which is set by the ratio of the self-propulsion velocity to the rotational diffusion constant. There have been numerous modeling and experimental studies of this minimal model system over the past five years, recently reviewed in \cite{Cates2015,marchetti2015structure,Bechinger2016}. Here we present a pedagogical overview of these results.

Although this material is composed of non-aligning particles with \textit{ no attractive interactions}, simulations showed that self-propelled colloids in 2D \cite{Fily2012,Redner2013,Stenhammar2013b,Buttinoni2013a,Mognetti2013} and 3D  \cite{Stenhammar2014b,Wysocki2013} phase separate into dense macroscopic clusters and a dilute phase (Fig.~\ref{fig:ActiveSpheres}A-C). Similar behavior was observed in experiments \cite{Buttinoni2013a}. The phase separation has all the hallmarks of equilibrium vapor-liquid phase coexistence. In equilibrium thermodynamics, the phase diagram of a vapor-liquid system as a function of density and temperature is described by the following terms: \textit{i)} a \textit{binodal}, which envelopes the region in paramater space where phase separation occurs (see Fig.~\ref{fig:ActiveSpheres}b); \textit{ii)} a \textit{spinodal}, which demarcates a region within the binodal where phase separation proceeds without a nucleation barrier, with clusters arising spontaneously throughout the fluid and then undergoing power-law coarsening; and \textit{iii)} a \textit{critical point} where the binodal and spinodal meet, at which density fluctuations become macroscopic in size. Active colloids exhibit all of these features \cite{Redner2013, Stenhammar2013b, Stenhammar2014b}. However, the usual role played by temperature in a gas-liquid system is now played by the inverse of the persistence length.

Phase separation  driven by particle self-propulsion is referred to as motility-induced phase separation (MIPS). Tailleur and Cates \cite{Tailleur2008,Cates2015} showed that particles whose propulsion velocity is a decreasing function of local particle density, \eg due to interparticle collisions, are generically unstable to phase separation. The decreasing velocity leads to local particle accumulation, and hence further velocity reduction, generating a positive feedback loop. Redner \etal \cite{Redner2013,Redner2013a} showed that the phase behavior of active colloids in 2D could be described by considering the fluxes of particles into/out of a macroscopic cluster (Fig.~\ref{fig:ActiveSpheres} D) The crucial observation is that a particle encountering a cluster surface is trapped by excluded volume interactions until its direction reorients away from the cluster. Since the inward flux of such particles is proportional to the density in the dilute phase, there is a threshold density above which the cluster phase will grow.  Both approaches show that the persistence of particle propulsion directions leads to \textit{self-trapping}, meaning a suppression of particle motions due to long-lived contacts between colliding particles. In summary, self-propulsion acts as an effective (albeit transient) interparticle attraction, driving emergent phenomena that passive hard spheres could never achieve.

While the gross bulk properties of self-propelled colloids are captured by the paradigm of gas-liquid phase separation, there are several surprising and unique properties:

1. The dense phase exhibits an unusual combination of structure and dynamics. When the particles are monodisperse, it exhibits the structural signatures of a liquid, hexatic, or crystal as the propulsion velocity increases, while polydisperse particles form active glasses \cite{Berthier2014, Ding2015}. Yet, even when the structure indicates a crystalline phase, the constituent particles are highly mobile, exhibiting diffusive behavior on long timescales \cite{Bialke2012, Redner2013}, more typical of a liquid. Moreover, mixtures of self-propelled and passive (Brownian) particles tend to segregate, with the active particles driving crystallization of their passive neighbors \cite{Ni2013a, Stenhammar2015}.
Additionally, while macrophase separation implies that active colloidal fluids have an effective surface tension, cluster boundaries exhibit large fluctuations which were found to be consistent with negative surface tension balanced by interfacial stiffness\cite{Bialke2015}.

2. Since self-propulsion acts as an effective attraction between repulsive colloids, one might expect that it would universally enhance coagulation of particles with attractive energetic interactions. However, self-propelled particles endowed with energetic interparticle attractions have a \textit{reentrant} phase diagram \cite{Redner2013a}; \ie, starting with an equilibrium system that is phase separated due to attractions, monotonically increasing self-propulsion first melts the system into a homogeneous fluid, and then returns it to a phase separated state. Extension of the model in Fig.~\ref{fig:ActiveSpheres}D describes this effect -- moderate propulsion strengths tend to break apart attraction-stabilized interactions, while strong propulsion leads to clustering via the self-trapping mechanism described above \cite{Redner2013a}. Importantly, the behavior at low propulsion strengths shows that the nonequilibrium properties of this system cannot simply be mapped onto an equilibrium system of particles with energetic attractions.

3. A pillar of materials theory is extensivity, which implies that boundary effects are local and studies of an isolated bulk system can be applied to real systems that are necessarily confined by boundaries. However, in active materials the effects of boundaries are profound and far reaching. For example, self-propelled colloids tend to accumulate at boundaries \cite{Elgeti2013,Yang2014b}, and are concentrated in regions of high curvature \cite{Fily2014b,Fily2015}. In another example, ratchets and funnels drive spontaneous flow in active fluids~\cite{Wan2008, Tailleur2009,Angelani2011,Ghosh2013,Ai2013}. This effect has been used to direct bacterial motion~\cite{Galajda2007} and harness bacterial power to propel microscopic gears~\cite{Angelani2009,DiLeonardo2010,Sokolov2010}.
%Thus, this simple realization of an active material has served as a guide in understanding the complexities that arise in this class of materials and has laid the theoretical foundations for understanding active matter in general and using this understanding to experimental systems of varying complexity.

\section*{Active Nematics}
We now consider active nematics, which consist of elongated particles whose activity enters with nematic (head-tail) symmetry. Biological active nematics include confluent monolayers of fibroblast cells \cite{Duclos2014} and growing cells that elongate over time \cite{Volfson2008}. The first synthetic active nematic system considered was that of a vibrated monolayer of granular rods \cite{Narayan2007}. Subsequently a robust, \textit{in vitro} biomaterial system composed of length-stabilized microtubules driven by motor protein clusters has served as a prototype active nematic \cite{Sanchez2012}. Synthetic and biological materials have also been combined, for example bacteria swimming in a lyotropic liquid crystal \cite{Zhou2014}.

\begin{figure*} [hbt]
\centering{\includegraphics[width=0.95\textwidth]{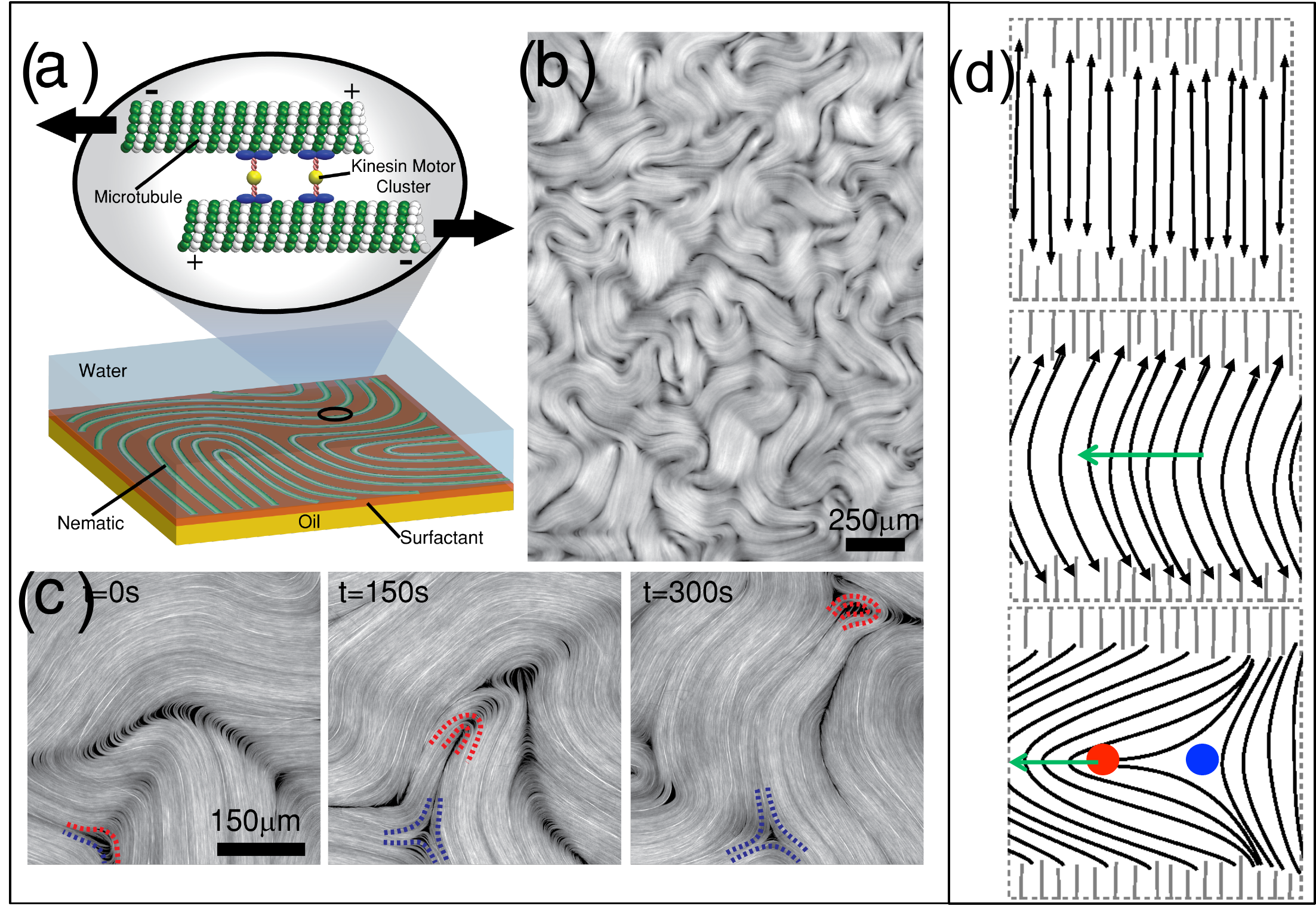}}
\caption{ \textbf{(a)-(c)} An experimental realization of an active nematic (provided by Stephen DeCamp). \textbf{(a)} A schematic of the experimental system that consists of length-stabilized microtubules and kinesin motor clusters that are depleted onto an air-water interface \cite{Sanchez2012,DeCamp2015}. \textbf{(b)} A zoomed out view of the experimental active nematic, showing a field of defects (localized discontinuities in the directions of the filaments). \textbf{(c)} A representative event in the experimental sample showing the creation and unbinding of a pair of defects. The defects have `charges' (measuring how the direction of filaments rotates around the defect) of $+1/2$ (outlined in red) and $-1/2$ (blue).  Forces are unbalanced within the $+1/2$ defect, propelling it across the sample.
 \textbf{(d)} Illustration of the primary mechanism for generation of defects in an active nematic. Top: Forces are balanced inside a region of perfect nematic order. The black arrows show the forces exerted by the extensile filaments on the background. Middle: When a bend fluctuation occurs, the forces become unbalanced, driving further growth of the fluctuation. The green arrow indicates the net force exerted by the background on the bending filaments. Bottom: Growth of the fluctuation leads to the nucleation and unbinding of a pair of $\pm 1/2$ defects. }
\label{fig:activenematics}
\end{figure*}

There is a vast theoretical literature on active nematics (recently reviewed in \cite{Marchetti2013,Ramaswamy2010}), beginning with seminal papers by Ramaswamy and collaborators \cite{AditiSimha2002,Mishra2006,Ramaswamy2003}. Diverse phenomenology have been uncovered and understood, including phase separation into unstable bands \cite{Ngo2014,Putzig2014a}, convection rolls and oscillatory patterns \cite{Giomi2011E,Marenduzzo2007}, spontaneous flow transitions due to confinement \cite{Marenduzzo2008,Giomi2008}, and anomalous rheological properties \cite{Hatwalne2004,Marenduzzo2007a,Saintillan2010,Cates2008,Giomi2010}. Here, we focus on the macroscale dynamics of dense 2D extensile systems (the particles push outward along their body axis), which is governed by topological defects.

The primary descriptor of a nematic fluid is the director field, which characterizes the local orientation of the constituent particles at any point in the fluid. A topological defect is a compact distortion of the director field that arises where domains of differently oriented nematic regions meet. Defects are characterized by a topological charge, which measures how much the director rotates around the defect. In nematics, the most common (and lowest charge) defects are $+1/2$ and $-1/2$ defects associated with the director field rotating by $180^\circ$ (Fig.~\ref{fig:activenematics}). In a passive nematic, defects are primarily generated during sample preparation, and are static. Nearby defects with opposite charge annihilate, lowering the elastic energy to approach the equilibrium state --- a uniformly aligned state with (quasi)-long-range orientational order.

The behavior of defects in active nematics is dramatically different. After sample preparation, defects continue to arise spontaneously until the system reaches a nonequilibrium steady-state with a finite defect density, with equal rates of defect generation and annihilation. Further, each $+1/2$ defect moves along the direction of its `head' as if self-propelled (Fig. [\ref{fig:activenematics}]). Powered by these motions, active nematics can exhibit large-scale turbulent flows, despite being in the low Reynolds number limit \cite{Wensink2012}, where turbulent flows typically do not occur because the fluid viscosity damps out inertial motions. In contrast to high Reynolds number turbulence, these flows exhibit an `inverse energy cascade', where energy injected at the particle scale generates large-scale flows.

These chaotic dynamics appear to destroy the (quasi)-long-range order characteristic of an equilibrium nematic. Remarkably, experiments on microtubule-based active nematics and simulations of extensile rods demonstrated that the defects themselves can acquire system-spanning orientational order \cite{DeCamp2015}. Defect-ordered phases were subsequently found in several theoretical models for active nematics \cite{doostmohammadi2015wet,oza2015nematic,Putzig2015}, suggesting they are a generic feature of active nematics. In contrast to equilibrium ordered phases of static topological defects (\eg twist grain boundaries in liquid crystals and flux line lattices in type II superconductors), the ordered defects in active nematics are motile and transient, continually undergoing annihilation and rebirth.

The mechanism of spontaneous defect generation and propulsion of the $+1/2$ defects can be understood as follows (Fig.~\ref{fig:activenematics}D). Extensile filaments exert forces along their long axis in both directions. These forces balance when the nematic is perfectly flat. However, a small fluctuation in local orientational order causes a net force orthogonal to the direction of mean order, driving growth of the fluctuation and eventual nucleation of a $\pm 1/2$ defect pair from the flat nematic background. Further, the distortion of the orientation field at a $+1/2$ defect generates a net force along its axis, which propels the defect along its `head'.

A number of computational and theoretical models have been developed, applicable to different experimental realizations of active nematics. These can be roughly categorized by whether they consider long-range fluid-mediated interactions (\cite{Thampi2013,Thampi2014,Thampi2014b,Giomi2013,Giomi2014}), or assume that fluid-mediated interactions are absent (such as vibrated rods or cell sheets) or screened at large scales due to confinement in 2D \cite{doostmohammadi2015wet,Pismen2013a,Putzig2015,Chate2006,Gao2014,Shi2013}. While all of these models capture spontaneous creation and self-propulsion of defects, interactions between defects are model-dependent, and which models most closely describe existing experimental systems requires further investigation.

\textbf{Outlook.}
The examples described above suggest that the symmetries of particle interactions and activity generation are a useful starting point for classifying the possible emergent behaviors in an active system. While inspiration and materials from biological systems were used to create new materials in these examples, researchers are also beginning to apply the physical understanding and theoretical frameworks developed for these materials to biology. For example, pattern formation in bacterial colonies has been described as  motility-induced phase separation \cite{Cates2010} that is arrested by the logistic growth of the bacteria, and the mitotic spindle in a zenopus egg extract has been quantitatively modeled as an active nematic fluid \cite{Brugues2014}. Thus, the paradigm of active matter has the potential to achieve biology-inspired advances in materials physics, while in turn shedding new light on the emergence of function in biological systems.

\begin{acknowledgements}

MFH and AB thank Stephen DeCamp for creating panels a-c of Fig.~\ref{fig:activenematics} and for providing the experimental images therein, and Gabriel Redner for assistance in creating the remaining figures. We acknowledge support from the Brandeis Center for Bioinspired Soft Materials, an NSF MRSEC,  DMR-1420382. AB also acknowledges support from NSF-DMR-1149266.

\end{acknowledgements}

%\bibliography{RednerDissertationWithAddtionsANDStars}

\begin{thebibliography}{10}
\expandafter\ifx\csname url\endcsname\relax
  \def\url#1{\texttt{#1}}\fi
\expandafter\ifx\csname urlprefix\endcsname\relax\def\urlprefix{URL }\fi
\expandafter\ifx\csname href\endcsname\relax
  \def\href#1#2{#2} \def\path#1{#1}\fi

\bibitem{Toner1995}
J.~Toner, Y.~Tu, {Long-Range Order in a Two-Dimensional Dynamical XY Model: How
  Birds Fly Together}, Phys. Rev. Lett. 75 (1995) 4326.
\newblock \href {http://dx.doi.org/10.1103/PhysRevLett.75.4326}
  {\path{doi:10.1103/PhysRevLett.75.4326}}.

\bibitem{Palacci2013a}
J.~Palacci, S.~Sacanna, A.~P. Steinberg, D.~J. Pine, P.~M. Chaikin, {Living
  Crystals of Light-Activated Colloidal Surfers}, Science 339~(6122) (2013)
  936, *Experiments on colloidal Janus particles that can be rendered
  sef-propelled through light activation show clustering behavior
  characteristic of active colloids.
\newblock \href {http://dx.doi.org/10.1126/science.1230020}
  {\path{doi:10.1126/science.1230020}}.

\bibitem{EColiImage2005}
\href{http://dx.doi.org/10.1371%2Fjournal.pbio.0030058}{Aging and death in e.
  coli}, PLoS Biol. 3~(2) (2005) e58.
\newblock \href {http://dx.doi.org/10.1371/journal.pbio.0030058}
  {\path{doi:10.1371/journal.pbio.0030058}}.
\newline\urlprefix\url{http://dx.doi.org/10.1371%2Fjournal.pbio.0030058}

\bibitem{Redner2013}
G.~S. Redner, M.~F. Hagan, A.~Baskaran, {Structure and Dynamics of a
  Phase-Separating Active Colloidal Fluid}, Phys. Rev. Lett. 110 (2013) 055701,
  * This work presents Brownian dynamics simulations and a kinetic theory
  describing motility induced phase separation in self-propelled colloids, and
  characterizes the dense phase of monodisperse colloids.
\newblock \href {http://dx.doi.org/10.1103/PhysRevLett.110.055701}
  {\path{doi:10.1103/PhysRevLett.110.055701}}.

\bibitem{Redner2013a}
G.~S. Redner, A.~Baskaran, M.~F. Hagan, {Reentrant phase behavior in active
  colloids with attraction}, Phys. Rev. E 88 (2013) 012305.
\newblock \href {http://dx.doi.org/10.1103/PhysRevE.88.012305}
  {\path{doi:10.1103/PhysRevE.88.012305}}.

\bibitem{Howse2007}
J.~R. Howse, R.~A.~L. Jones, A.~J. Ryan, T.~Gough, R.~Vafabakhsh,
  R.~Golestanian, {Self-Motile Colloidal Particles: From Directed Propulsion to
  Random Walk}, Phys. Rev. Lett. 99 (2007) 048102.
\newblock \href {http://dx.doi.org/10.1103/PhysRevLett.99.048102}
  {\path{doi:10.1103/PhysRevLett.99.048102}}.

\bibitem{Ke2010}
H.~Ke, S.~Ye, R.~L. Carroll, K.~Showalter, {Motion analysis of self-propelled
  Pt-silica particles in hydrogen peroxide solutions.}, J. Phys. Chem. A
  114~(17) (2010) 5462.
\newblock \href {http://dx.doi.org/10.1021/jp101193u}
  {\path{doi:10.1021/jp101193u}}.

\bibitem{Theurkauff2012a}
I.~Theurkauff, C.~Cottin-Bizonne, J.~Palacci, C.~Ybert, L.~Bocquet, {Dynamic
  Clustering in Active Colloidal Suspensions with Chemical Signaling}, Phys.
  Rev. Lett. 108 (2012) 268303.
\newblock \href {http://dx.doi.org/10.1103/PhysRevLett.108.268303}
  {\path{doi:10.1103/PhysRevLett.108.268303}}.

\bibitem{Palacci2014}
J.~Palacci, S.~Sacanna, S.-H. Kim, G.-R. Yi, D.~J. Pine, P.~M. Chaikin,
  {Light-activated self-propelled colloids}, Phil. Trans. R. Soc. A 372~(2029)
  (2014) 20130372.
\newblock \href {http://dx.doi.org/10.1098/rsta.2013.0372}
  {\path{doi:10.1098/rsta.2013.0372}}.

\bibitem{Buttinoni2013a}
I.~Buttinoni, J.~Bialk\'{e}, F.~K\"{u}mmel, H.~L\"{o}wen, C.~Bechinger,
  T.~Speck, {Dynamical Clustering and Phase Separation in Suspensions of
  Self-Propelled Colloidal Particles}, Phys. Rev. Lett. 110 (2013) 238301, *
  This paper shows that an experimental system of diffusophoretic colloids and
  Brownian dynamics simulations of self-propelled spheres with no aligning
  interactions both exhibit motility-induced phase separation.
\newblock \href {http://dx.doi.org/10.1103/PhysRevLett.110.238301}
  {\path{doi:10.1103/PhysRevLett.110.238301}}.

\bibitem{Cates2015}
M.~E. Cates, J.~Tailleur,
  \href{http://dx.doi.org/10.1146/annurev-conmatphys-031214-014710}{Motility-induced
  phase separation}, Annu. Rev. Condens. Matter Phys. 6~(1) (2015) 219--244, *
  Recent comprehensive review on the phenomenology and theory of motility
  induced phase separation by experts in the field.
\newblock \href
  {http://arxiv.org/abs/http://dx.doi.org/10.1146/annurev-conmatphys-031214-014710}
  {\path{arXiv:http://dx.doi.org/10.1146/annurev-conmatphys-031214-014710}},
  \href {http://dx.doi.org/10.1146/annurev-conmatphys-031214-014710}
  {\path{doi:10.1146/annurev-conmatphys-031214-014710}}.
\newline\urlprefix\url{http://dx.doi.org/10.1146/annurev-conmatphys-031214-014710}

\bibitem{marchetti2015structure}
M.~C. Marchetti, Y.~Fily, S.~Henkes, A.~Patch, D.~Yllanes, Structure and
  mechanics of active colloids, arXiv preprint arXiv:1510.00425.

\bibitem{Bechinger2016}
C.~Bechinger, R.~D. Leonardo, H.~Lowen, C.~Reichhardt, G.~Volpe, G.~Volpe,
  \href{http://arXiv.org/abs/1602.00081}{Active brownian particles in complex
  and crowded environments}, arXiv:1602.00081.
\newline\urlprefix\url{http://arXiv.org/abs/1602.00081}

\bibitem{Fily2012}
Y.~Fily, M.~C. Marchetti, {Athermal Phase Separation of Self-Propelled
  Particles with No Alignment}, Phys. Rev. Lett. 108 (2012) 235702, * The first
  account of motility induced phase separation in self-propelled colloids
  studied by Brownian dynamics simulations.
\newblock \href {http://dx.doi.org/10.1103/PhysRevLett.108.235702}
  {\path{doi:10.1103/PhysRevLett.108.235702}}.

\bibitem{Stenhammar2013b}
J.~Stenhammar, A.~Tiribocchi, R.~J. Allen, D.~Marenduzzo, M.~E. Cates,
  {Continuum theory of phase separation kinetics for active brownian
  particles}, Phys. Rev. Lett. 111 (2013) 145702, * This paper derives an
  effective continuum theory for motility-induced phase separation of particles
  with repulsive interactions, which has the standard bulk free energy for
  particles at equilibrium phase coexistence, but additional gradient terms
  which violate detailed balance.
\newblock \href {http://dx.doi.org/10.1103/PhysRevLett.111.145702}
  {\path{doi:10.1103/PhysRevLett.111.145702}}.

\bibitem{Mognetti2013}
B.~M. Mognetti, S.~A., S.~Angioletti-Uberti, A.~Cacciuto, C.~Valeriani,
  D.~Frenkel,
  \href{http://link.aps.org/doi/10.1103/PhysRevLett.111.245702}{Living clusters
  and crystals from low-density suspensions of active colloids}, Phys. Rev.
  Lett. 111 (2013) 245702.
\newblock \href {http://dx.doi.org/10.1103/PhysRevLett.111.245702}
  {\path{doi:10.1103/PhysRevLett.111.245702}}.
\newline\urlprefix\url{http://link.aps.org/doi/10.1103/PhysRevLett.111.245702}

\bibitem{Stenhammar2014b}
J.~Stenhammar, D.~Marenduzzo, R.~J. Allen, M.~E. Cates, {Phase behaviour of
  active Brownian particles: the role of dimensionality.}, Soft Matter 10~(10)
  (2014) 1489.
\newblock \href {http://dx.doi.org/10.1039/c3sm52813h}
  {\path{doi:10.1039/c3sm52813h}}.

\bibitem{Wysocki2013}
A.~Wysocki, R.~G. Winkler, G.~Gompper,
  \href{http://arxiv.org/abs/1308.6423v1}{Cooperative motion of active brownian
  spheres in three-dimensional dense suspensions} (2013).
\newline\urlprefix\url{http://arxiv.org/abs/1308.6423v1}

\bibitem{Tailleur2008}
J.~Tailleur, M.~Cates, {Statistical Mechanics of Interacting Run-and-Tumble
  Bacteria}, Phys. Rev. Lett. 100 (2008) 218103.
\newblock \href {http://dx.doi.org/10.1103/PhysRevLett.100.218103}
  {\path{doi:10.1103/PhysRevLett.100.218103}}.

\bibitem{Berthier2014}
L.~Berthier, {Nonequilibrium Glassy Dynamics of Self-Propelled Hard Disks},
  Phys. Rev. Lett. 112 (2014) 220602.
\newblock \href {http://dx.doi.org/10.1103/PhysRevLett.112.220602}
  {\path{doi:10.1103/PhysRevLett.112.220602}}.

\bibitem{Ding2015}
H.~Ding, M.~Feng, H.~Jiang, Z.~Hou, {Nonequilibrium Glass Transition in
  Mixtures of Active-Passive Particles}\href
  {http://arxiv.org/abs/arXiv:1506.02754} {\path{arXiv:arXiv:1506.02754}}.

\bibitem{Bialke2012}
J.~Bialk\'e, T.~Speck, H.~L\"owen, Crystallization in a dense suspension of
  self-propelled particles, Phys. Rev. Lett. 108 (2012) 168301.
\newblock \href {http://dx.doi.org/10.1103/PhysRevLett.108.168301}
  {\path{doi:10.1103/PhysRevLett.108.168301}}.

\bibitem{Ni2013a}
R.~Ni, M.~a.~C. Stuart, M.~Dijkstra, P.~G. Bolhuis, {Crystallizing hard-sphere
  glasses by doping with active particles}, Soft Matter 10 (2013) 6609.
\newblock \href {http://dx.doi.org/10.1039/C4SM01015A}
  {\path{doi:10.1039/C4SM01015A}}.

\bibitem{Stenhammar2015}
J.~Stenhammar, R.~Wittkowski, D.~Marenduzzo, M.~E. Cates, {Activity-Induced
  Phase Separation and Self-Assembly in Mixtures of Active and Passive
  Particles}, Phys. Rev. Lett. 114 (2015) 018301.
\newblock \href {http://dx.doi.org/10.1103/PhysRevLett.114.018301}
  {\path{doi:10.1103/PhysRevLett.114.018301}}.

\bibitem{Bialke2015}
J.~Bialk\'{e}, J.~T. Siebert, H.~L\"{o}wen, T.~Speck, {Negative Interfacial
  Tension in Phase-Separated Active Brownian Particles}, Phys. Rev. Lett. 115
  (2015) 098301.
\newblock \href {http://dx.doi.org/10.1103/PhysRevLett.115.098301}
  {\path{doi:10.1103/PhysRevLett.115.098301}}.

\bibitem{Elgeti2013}
J.~Elgeti, G.~Gompper,
  \href{http://stacks.iop.org/0295-5075/101/i=4/a=48003}{Wall accumulation of
  self-propelled spheres}, Europhys. Lett. 101~(4) (2013) 48003.
\newline\urlprefix\url{http://stacks.iop.org/0295-5075/101/i=4/a=48003}

\bibitem{Yang2014b}
X.~Yang, M.~L. Manning, M.~C. Marchetti,
  \href{http://dx.doi.org/10.1039/C4SM00927D}{Aggregation and segregation of
  confined active particles}, Soft Matter 10 (2014) 6477--6484.
\newblock \href {http://dx.doi.org/10.1039/C4SM00927D}
  {\path{doi:10.1039/C4SM00927D}}.
\newline\urlprefix\url{http://dx.doi.org/10.1039/C4SM00927D}

\bibitem{Fily2014b}
Y.~Fily, A.~Baskaran, M.~F. Hagan, {Dynamics of self-propelled particles under
  strong confinement.}, Soft Matter 10~(30) (2014) 5609.
\newblock \href {http://dx.doi.org/10.1039/c4sm00975d}
  {\path{doi:10.1039/c4sm00975d}}.

\bibitem{Fily2015}
Y.~Fily, A.~Baskaran, M.~F. Hagan, {Dynamics and density distribution of
  strongly confined noninteracting nonaligning self-propelled particles in a
  nonconvex boundary}, Phys. Rev. E 91 (2015) 012125.
\newblock \href {http://dx.doi.org/10.1103/PhysRevE.91.012125}
  {\path{doi:10.1103/PhysRevE.91.012125}}.

\bibitem{Wan2008}
M.~B. Wan, C.~J. Olson~Reichhardt, Z.~Nussinov, C.~Reichhardt, Rectification of
  swimming bacteria and self-driven particle systems by arrays of asymmetric
  barriers, Phys. Rev. Lett. 101~(1) (2008) 018102.
\newblock \href {http://dx.doi.org/10.1103/PhysRevLett.101.018102}
  {\path{doi:10.1103/PhysRevLett.101.018102}}.

\bibitem{Tailleur2009}
J.~Tailleur, M.~E. Cates,
  \href{http://stacks.iop.org/0295-5075/86/i=6/a=60002}{Sedimentation,
  trapping, and rectification of dilute bacteria}, Europhys. Lett. 86~(6)
  (2009) 60002.
\newline\urlprefix\url{http://stacks.iop.org/0295-5075/86/i=6/a=60002}

\bibitem{Angelani2011}
L.~Angelani, A.~Costanzo, R.~D. Leonardo, Active ratchets, Europhys. Lett.
  96~(6) (2011) 68002.
\newblock \href {http://dx.doi.org/10.1209/0295-5075/96/68002}
  {\path{doi:10.1209/0295-5075/96/68002}}.

\bibitem{Ghosh2013}
P.~K. Ghosh, V.~R. Misko, F.~Marchesoni, F.~Nori, Self-propelled janus
  particles in a ratchet: Numerical simulations, Phys. Rev. Lett. 110~(26)
  (2013) 268301.
\newblock \href {http://dx.doi.org/10.1103/PhysRevLett.110.268301}
  {\path{doi:10.1103/PhysRevLett.110.268301}}.

\bibitem{Ai2013}
B.-q. Ai, Q.-y. Chen, Y.-f. He, F.-g. Li, W.-r. Zhong, Rectification and
  diffusion of self-propelled particles in a two-dimensional corrugated
  channel, Phys. Rev. E 88~(6) (2013) 062129.
\newblock \href {http://dx.doi.org/10.1103/PhysRevE.88.062129}
  {\path{doi:10.1103/PhysRevE.88.062129}}.

\bibitem{Galajda2007}
P.~Galajda, J.~Keymer, P.~Chaikin, R.~Austin, {A Wall of Funnels Concentrates
  Swimming Bacteria}, J. Bacteriol. 189~(23) (2007) 8704.
\newblock \href {http://dx.doi.org/10.1128/JB.01033-07}
  {\path{doi:10.1128/JB.01033-07}}.

\bibitem{Angelani2009}
L.~Angelani, R.~Di~Leonardo, G.~Ruocco, Self-starting micromotors in a
  bacterial bath, Phys. Rev. Lett. 102~(4) (2009) 048104.
\newblock \href {http://dx.doi.org/10.1103/PhysRevLett.102.048104}
  {\path{doi:10.1103/PhysRevLett.102.048104}}.

\bibitem{DiLeonardo2010}
R.~{Di Leonardo}, L.~Angelani, D.~Dell'Arciprete, G.~Ruocco, V.~Iebba,
  S.~Schippa, M.~P. Conte, F.~Mecarini, F.~{De Angelis}, E.~{Di Fabrizio},
  {Bacterial ratchet motors}, Proc. Natl. Acad. Sci. U. S. A. 107~(21) (2010)
  9541.
\newblock \href {http://dx.doi.org/10.1073/pnas.0910426107}
  {\path{doi:10.1073/pnas.0910426107}}.

\bibitem{Sokolov2010}
A.~Sokolov, M.~M. Apodaca, B.~A. Grzybowski, I.~S. Aranson, {Swimming bacteria
  power microscopic gears}, Proc. Natl. Acad. Sci. U. S. A. 107~(3) (2010) 969.
\newblock \href {http://dx.doi.org/10.1073/pnas.0913015107}
  {\path{doi:10.1073/pnas.0913015107}}.

\bibitem{Duclos2014}
G.~Duclos, S.~Garcia, H.~G. Yevick, P.~Silberzan, {Perfect nematic order in
  confined monolayers of spindle-shaped cells}, Soft Matter 10~(14) (2014)
  2346.
\newblock \href {http://dx.doi.org/10.1039/c3sm52323c}
  {\path{doi:10.1039/c3sm52323c}}.

\bibitem{Volfson2008}
D.~Volfson, S.~Cookson, J.~Hasty, L.~S. Tsimring, {Biomechanical ordering of
  dense cell populations.}, Proc. Natl. Acad. Sci. U. S. A. 105~(40) (2008)
  15346.
\newblock \href {http://dx.doi.org/10.1073/pnas.0706805105}
  {\path{doi:10.1073/pnas.0706805105}}.

\bibitem{Narayan2007}
V.~Narayan, S.~Ramaswamy, N.~Menon, {Long-lived giant number fluctuations in a
  swarming granular nematic}, Science 317~(5834) (2007) 105.
\newblock \href {http://dx.doi.org/10.1126/science.1140414}
  {\path{doi:10.1126/science.1140414}}.

\bibitem{Sanchez2012}
T.~Sanchez, D.~T.~N. Chen, S.~J. DeCamp, M.~Heymann, Z.~Dogic, {Spontaneous
  motion in hierarchically assembled active matter}, Nature 491~(7424) (2012)
  431.
\newblock \href {http://dx.doi.org/10.1038/nature11591}
  {\path{doi:10.1038/nature11591}}.

\bibitem{Zhou2014}
S.~Zhou, A.~Sokolov, O.~D. Lavrentovich, I.~S. Aranson, {Living liquid
  crystals}, Proc. Natl. Acad. Sci. U. S. A. 111~(4) (2014) 1265.
\newblock \href {http://dx.doi.org/10.1073/pnas.1321926111}
  {\path{doi:10.1073/pnas.1321926111}}.

\bibitem{Marchetti2013}
M.~C. Marchetti, J.~F. Joanny, S.~Ramaswamy, T.~B. Liverpool, J.~Prost, M.~Rao,
  R.~A. Simha, {Hydrodynamics of soft active matter}, Reviews of Modern Physics
  85 (2013) 1143, * An extensive review of the active liquid crystal literature
  by the experts in the field.
\newblock \href {http://dx.doi.org/10.1103/RevModPhys.85.1143}
  {\path{doi:10.1103/RevModPhys.85.1143}}.

\bibitem{Ramaswamy2010}
S.~Ramaswamy, The mechanics and statistics of active matter, Annu. Rev.
  Condens. Matter Phys. 1 (2010) 323--345.

\bibitem{AditiSimha2002}
R.~{Aditi Simha}, S.~Ramaswamy, {Hydrodynamic Fluctuations and Instabilities in
  Ordered Suspensions of Self-Propelled Particles}, Phys. Rev. Lett. 89 (2002)
  058101.
\newblock \href {http://dx.doi.org/10.1103/PhysRevLett.89.058101}
  {\path{doi:10.1103/PhysRevLett.89.058101}}.

\bibitem{Mishra2006}
S.~Mishra, S.~Ramaswamy, {Active Nematics Are Intrinsically Phase Separated},
  Phys. Rev. Lett. 97 (2006) 090602.
\newblock \href {http://dx.doi.org/10.1103/PhysRevLett.97.090602}
  {\path{doi:10.1103/PhysRevLett.97.090602}}.

\bibitem{Ramaswamy2003}
S.~Ramaswamy, R.~A. Simha, J.~Toner, {Active nematics on a substrate: Giant
  number fluctuations and long-time tails}, Europhys. Lett. 62~(2) (2003) 196.
\newblock \href {http://dx.doi.org/10.1209/epl/i2003-00346-7}
  {\path{doi:10.1209/epl/i2003-00346-7}}.

\bibitem{Ngo2014}
S.~Ngo, A.~Peshkov, I.~S. Aranson, E.~Bertin, F.~Ginelli, H.~Chat\'{e},
  {Large-Scale Chaos and Fluctuations in Active Nematics}, Phys. Rev. Lett. 113
  (2014) 038302.
\newblock \href {http://dx.doi.org/10.1103/PhysRevLett.113.038302}
  {\path{doi:10.1103/PhysRevLett.113.038302}}.

\bibitem{Putzig2014a}
E.~Putzig, A.~Baskaran, {Phase separation and emergent structures in an active
  nematic fluid}, Phys. Rev. E 90 (2014) 042304.
\newblock \href {http://dx.doi.org/10.1103/PhysRevE.90.042304}
  {\path{doi:10.1103/PhysRevE.90.042304}}.

\bibitem{Giomi2011E}
L.~Giomi, L.~Mahadevan, B.~Chakraborty, M.~F. Hagan, Excitable patterns in
  active nematics, Phys. Rev. Lett. 106 (2011) 218101.
\newblock \href {http://dx.doi.org/10.1103/PhysRevLett.106.218101}
  {\path{doi:10.1103/PhysRevLett.106.218101}}.

\bibitem{Marenduzzo2007}
D.~Marenduzzo, E.~Orlandini, M.~Cates, J.~Yeomans, {Steady-state hydrodynamic
  instabilities of active liquid crystals: Hybrid lattice Boltzmann
  simulations}, Phys. Rev. E 76 (2007) 031921.
\newblock \href {http://dx.doi.org/10.1103/PhysRevE.76.031921}
  {\path{doi:10.1103/PhysRevE.76.031921}}.

\bibitem{Marenduzzo2008}
D.~Marenduzzo, E.~Orlandini, M.~Cates, J.~Yeomans, {Lattice Boltzmann
  simulations of spontaneous flow in active liquid crystals: The role of
  boundary conditions}, J. Non-Newtonian Fluid Mech. 149~(1-3) (2008) 56.
\newblock \href {http://dx.doi.org/10.1016/j.jnnfm.2007.02.005}
  {\path{doi:10.1016/j.jnnfm.2007.02.005}}.

\bibitem{Giomi2008}
L.~Giomi, M.~Marchetti, T.~Liverpool, {Complex Spontaneous Flows and
  Concentration Banding in Active Polar Films}, Phys. Rev. Lett. 101 (2008)
  198101.
\newblock \href {http://dx.doi.org/10.1103/PhysRevLett.101.198101}
  {\path{doi:10.1103/PhysRevLett.101.198101}}.

\bibitem{Hatwalne2004}
Y.~Hatwalne, S.~Ramaswamy, M.~Rao, R.~Simha, {Rheology of Active-Particle
  Suspensions}, Phys. Rev. Lett. 92 (2004) 118101.
\newblock \href {http://dx.doi.org/10.1103/PhysRevLett.92.118101}
  {\path{doi:10.1103/PhysRevLett.92.118101}}.

\bibitem{Marenduzzo2007a}
D.~Marenduzzo, E.~Orlandini, J.~Yeomans, {Hydrodynamics and Rheology of Active
  Liquid Crystals: A Numerical Investigation}, Phys. Rev. Lett. 98 (2007)
  118102.
\newblock \href {http://dx.doi.org/10.1103/PhysRevLett.98.118102}
  {\path{doi:10.1103/PhysRevLett.98.118102}}.

\bibitem{Saintillan2010}
D.~Saintillan, {Extensional rheology of active suspensions}, Phys. Rev. E 81
  (2010) 056307.
\newblock \href {http://dx.doi.org/10.1103/PhysRevE.81.056307}
  {\path{doi:10.1103/PhysRevE.81.056307}}.

\bibitem{Cates2008}
M.~Cates, S.~Fielding, D.~Marenduzzo, E.~Orlandini, J.~Yeomans, {Shearing
  Active Gels Close to the Isotropic-Nematic Transition}, Phys. Rev. Lett. 101
  (2008) 068102.
\newblock \href {http://dx.doi.org/10.1103/PhysRevLett.101.068102}
  {\path{doi:10.1103/PhysRevLett.101.068102}}.

\bibitem{Giomi2010}
L.~Giomi, T.~B. Liverpool, M.~C. Marchetti, {Sheared active fluids: Thickening,
  thinning, and vanishing viscosity}, Phys. Rev. E 81 (2010) 051908.
\newblock \href {http://dx.doi.org/10.1103/PhysRevE.81.051908}
  {\path{doi:10.1103/PhysRevE.81.051908}}.

\bibitem{Wensink2012}
H.~H. Wensink, J.~Dunkel, S.~Heidenreich, K.~Drescher, R.~E. Goldstein,
  H.~Lowen, J.~M. Yeomans, {Meso-scale turbulence in living fluids}, Proc.
  Natl. Acad. Sci. U. S. A. 109~(36) (2012) 14308--14313, * This article
  combines experiments, particle-based simulations, and continuum hydrodynamic
  theory to characterize turbulence arising in active matter systems from the
  injection of energy at the microscale.
\newblock \href {http://dx.doi.org/10.1073/pnas.1202032109}
  {\path{doi:10.1073/pnas.1202032109}}.

\bibitem{DeCamp2015}
S.~J. DeCamp, G.~S. Redner, A.~Baskaran, M.~F. Hagan, Z.~Dogic, {Orientational
  order of motile defects in active nematics}, Nat. Mater. 32 (2015) 1, *
  Identifying and tracking defects within experimental and computational
  realizations of active nematics showed that, for certain parameter ranges,
  topological defects acquire long-range order.
\newblock \href {http://dx.doi.org/10.1038/nmat4387}
  {\path{doi:10.1038/nmat4387}}.

\bibitem{doostmohammadi2015wet}
A.~Doostmohammadi, M.~Adamer, S.~P. Thampi, J.~M. Yeomans, Wet to dry crossover
  and a flow vortex-lattice in active nematics, arXiv preprint
  arXiv:1505.04199.

\bibitem{oza2015nematic}
A.~U. Oza, J.~Dunkel, Nematic ordering of topological defects in active liquid
  crystals, arXiv preprint arXiv:1507.01055.

\bibitem{Putzig2015}
E.~Putzig, G.~S. Redner, A.~Baskaran, A.~Baskaran, {Instabilities, defects, and
  defect ordering in an overdamped active nematic}\href
  {http://arxiv.org/abs/arXiv:1506.03501v1} {\path{arXiv:arXiv:1506.03501v1}}.

\bibitem{Thampi2013}
S.~P. Thampi, R.~Golestanian, J.~M. Yeomans, {Velocity Correlations in an
  Active Nematic}, Phys. Rev. Lett 111~(11) (2013) 118101.
\newblock \href {http://dx.doi.org/10.1103/PhysRevLett.111.118101}
  {\path{doi:10.1103/PhysRevLett.111.118101}}.

\bibitem{Thampi2014}
S.~P. Thampi, R.~Golestanian, J.~M. Yeomans, Vorticity, defects and
  correlations in active turbulence, Phil. Trans. R. Soc. A 372~(2029) (2014)
  20130366.

\bibitem{Thampi2014b}
S.~P. Thampi, R.~Golestanian, J.~M. Yeomans, {Instabilities and topological
  defects in active nematics}, Europhys. Lett. 105~(1) (2014) 18001.
\newblock \href {http://dx.doi.org/10.1209/0295-5075/105/18001}
  {\path{doi:10.1209/0295-5075/105/18001}}.

\bibitem{Giomi2013}
L.~Giomi, N.~Hawley-Weld, L.~Mahadevan, {Swarming, swirling and stasis in
  sequestered bristle-bots}~(Turner 2011) (2013) 21.
\newblock \href {http://dx.doi.org/10.1098/rspa.2012.0637}
  {\path{doi:10.1098/rspa.2012.0637}}.

\bibitem{Giomi2014}
L.~Giomi, M.~J. Bowick, P.~Mishra, R.~Sknepnek, M.~C. Marchetti, Defect
  dynamics in active nematics, Phil. Trans. R. Soc. A 372~(2029) (2014)
  20130365.

\bibitem{Pismen2013a}
L.~M. Pismen, {Dynamics of defects in an active nematic layer}, Phys. Rev. E
  88~(5) (2013) 050502.
\newblock \href {http://dx.doi.org/10.1103/PhysRevE.88.050502}
  {\path{doi:10.1103/PhysRevE.88.050502}}.

\bibitem{Chate2006}
H.~Chat\'e, F.~Ginelli, R.~Montagne, Simple model for active nematics:
  Quasi-long-range order and giant fluctuations, Phys. Rev. Lett. 96~(18)
  (2006) 180602.
\newblock \href {http://dx.doi.org/10.1103/PhysRevLett.96.180602}
  {\path{doi:10.1103/PhysRevLett.96.180602}}.

\bibitem{Gao2014}
T.~Gao, R.~Blackwell, M.~A. Glaser, M.~Betterton, M.~J. Shelley, Multiscale
  polar theory of microtubule and motor-protein assemblies, Phys. Rev. Lett.
  114 (2015) 048101.
\newblock \href {http://dx.doi.org/10.1103/PhysRevLett.114.048101}
  {\path{doi:10.1103/PhysRevLett.114.048101}}.

\bibitem{Shi2013}
X.~Q. Shi, Y.~Q. Ma, {Topological structure dynamics revealing collective
  evolution in active nematics}, Nat. Comm. 4 (2013) 3013.
\newblock \href {http://dx.doi.org/10.1038/ncomms4013}
  {\path{doi:10.1038/ncomms4013}}.

\bibitem{Cates2010}
M.~E. Cates, D.~Marenduzzo, I.~Pagonabarraga, J.~Tailleur, {Arrested phase
  separation in reproducing bacteria creates a generic route to pattern
  formation}, Proc. Natl. Acad. Sci. USA 107~(26) (2010) 11715.
\newblock \href {http://dx.doi.org/10.1073/pnas.1001994107}
  {\path{doi:10.1073/pnas.1001994107}}.

\bibitem{Brugues2014}
J.~Brugues, D.~Needleman, Physical basis of spindle self-organization,
  Proceedings of the National Academy of Sciences 111~(52) (2014) 18496--18500,
  * Extensive analysis of spindles assembled within cellular extract shows that
  many of their properties can be described by symmetry-based hydrodynamic
  theory, and elucidates the mechanisms that control the spindle size and
  shape.
\newblock \href {http://dx.doi.org/10.1073/pnas.1409404111}
  {\path{doi:10.1073/pnas.1409404111}}.

\end{thebibliography}

\end{document}